\documentclass[iop]{emulateapj}
\usepackage{color}
\usepackage{lscape}
\newcommand{\HI}{\ion{H}{1}~}
\newcommand{\Lya}{Lyman~$\alpha$~}
\newcommand{\kms}{km~s$^{-1}$~}

\shorttitle{Small-scale Properties of HI in Extended Disks}
\shortauthors{Borthakur et al.}

\begin{document}

\title{Small-scale Properties of Atomic Gas in Extended Disks of Galaxies}

\author{Sanchayeeta Borthakur}
\affil{Department of Physics and Astronomy, Johns Hopkins University, Baltimore, MD 21218, USA}
\email{sanch@pha.jhu.edu}

\author{Emmanuel Momjian}
\affil{National Radio Astronomy Observatory, Socorro, NM, USA}

\author{Timothy M. Heckman}
\affil{Department of Physics and Astronomy, Johns Hopkins University, Baltimore, MD 21218, USA}

\author{Donald G. York }
\affil{Department of Astronomy and Astrophysics, University of Chicago, Chicago, IL 60637, USA; Enrico Fermi Institute, University of Chicago, Chicago, IL 60637, USA}

\author{David V. Bowen }
\affil{Princeton University Observatory, Peyton Hall, Ivy Lane, Princeton NJ 08544}

\author{Min S. Yun}
\affil{Astronomy Department, University of Massachusetts, Amherst, MA 01003, USA}

\author{Todd M. Tripp}
\affil{Department of Astronomy, University of Massachusetts, Amherst, MA 01003, USA}

\begin{abstract}

We present high-resolution \HI 21~cm observations with the Karl G. Jansky Very Large Array (VLA) for three \HI rich galaxies in absorption against radio quasars.
Our sample contains six sightlines with impact parameters from 2.6 to 32.4~kpc. 
We detected a narrow \HI\ absorber of FWHM 1.1~\kms at 444.5~\kms towards  J122106.854+454852.16 probing the dwarf galaxy UCG~7408 at an impact parameter of 2.8~kpc. The absorption feature was barely resolved and its width corresponds to a maximum kinetic temperature, $\rm T_k \approx 26~K$.
We estimate a limiting peak optical depth of 1.37 and a column density of $\rm 6\times 10^{19}~cm^{-2}$. The physical extent of the absorber is $\rm 0.04~kpc^2$ and covers $\sim$25-30\% of the background source. 
A comparison between the emission and absorption strengths suggests the cold-to-total \HI column density in the absorber is $\sim$30\%. Folding in the covering fraction, the cold-to-total \HI mass is $\sim10\%$. This suggest that condensation of warm \HI ($\rm T_s\sim 1000~K$) to cold phase ($\rm T_s < 100~K$) is suppressed in UGC~7408. The unusually low temperature of the \HI absorber also indicates inefficiency in condensation of atomic  gas into molecular gas. 
The suppression in condensation is likely to be the result of low-metal content in this galaxy. The same process might explain the low efficiency of star formation in dwarf galaxies despite their huge of gas reservoirs.
We also report the non-detection of \HI\ in absorption in five other sightlines. This indicates that either the cold gas distribution is highly patchy or the gas is much warmer ($\rm T_s~>1000~K$) towards these sightlines.

\end{abstract}


\section{INTRODUCTION \label{Sec:intro}}

\begin{figure*}
\figurenum{1} 
\includegraphics[trim= 0mm 0mm 0mm 0mm, clip=true, scale=0.46,angle=0]{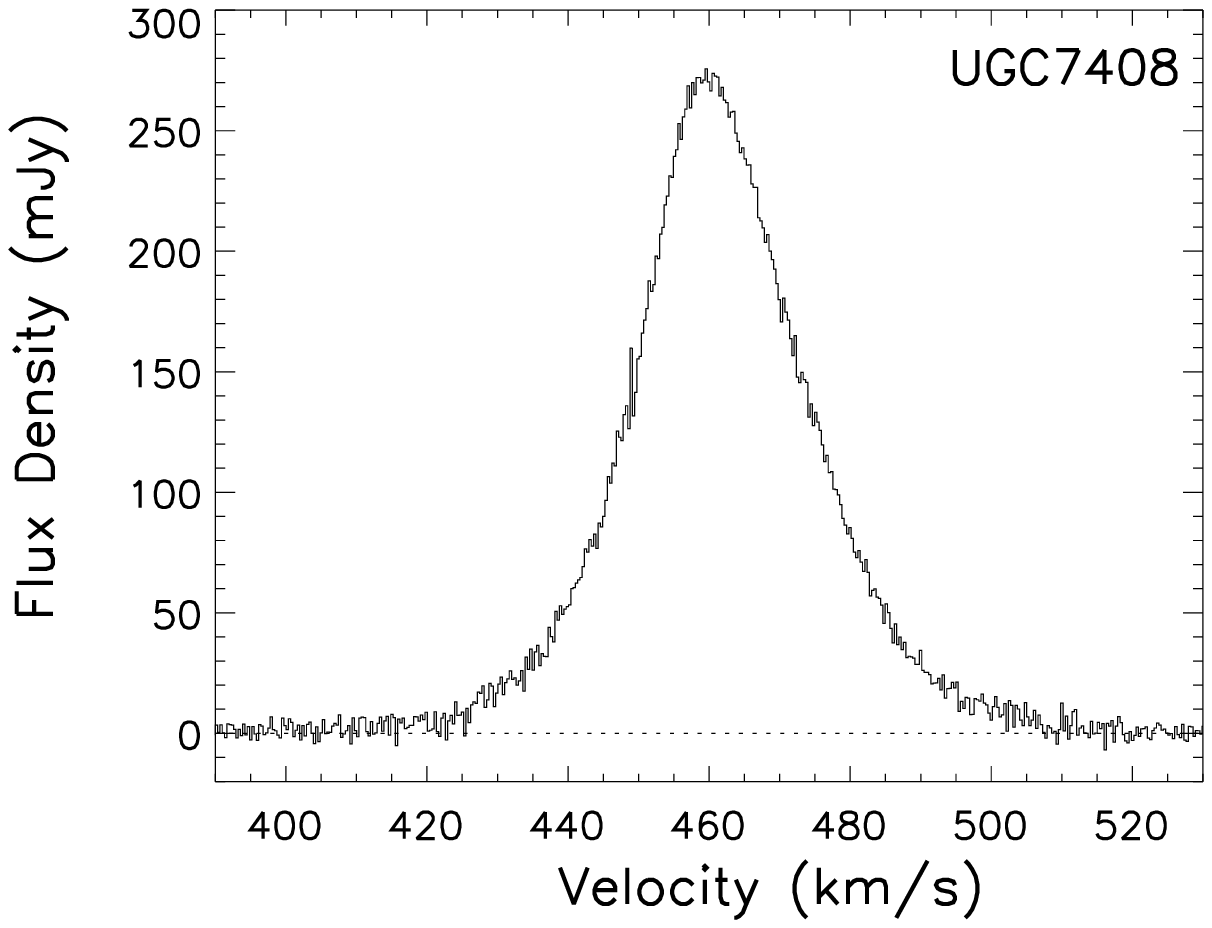} 
\includegraphics[trim= 0mm 0mm 0mm 0mm, clip=true, scale=0.46,angle=0]{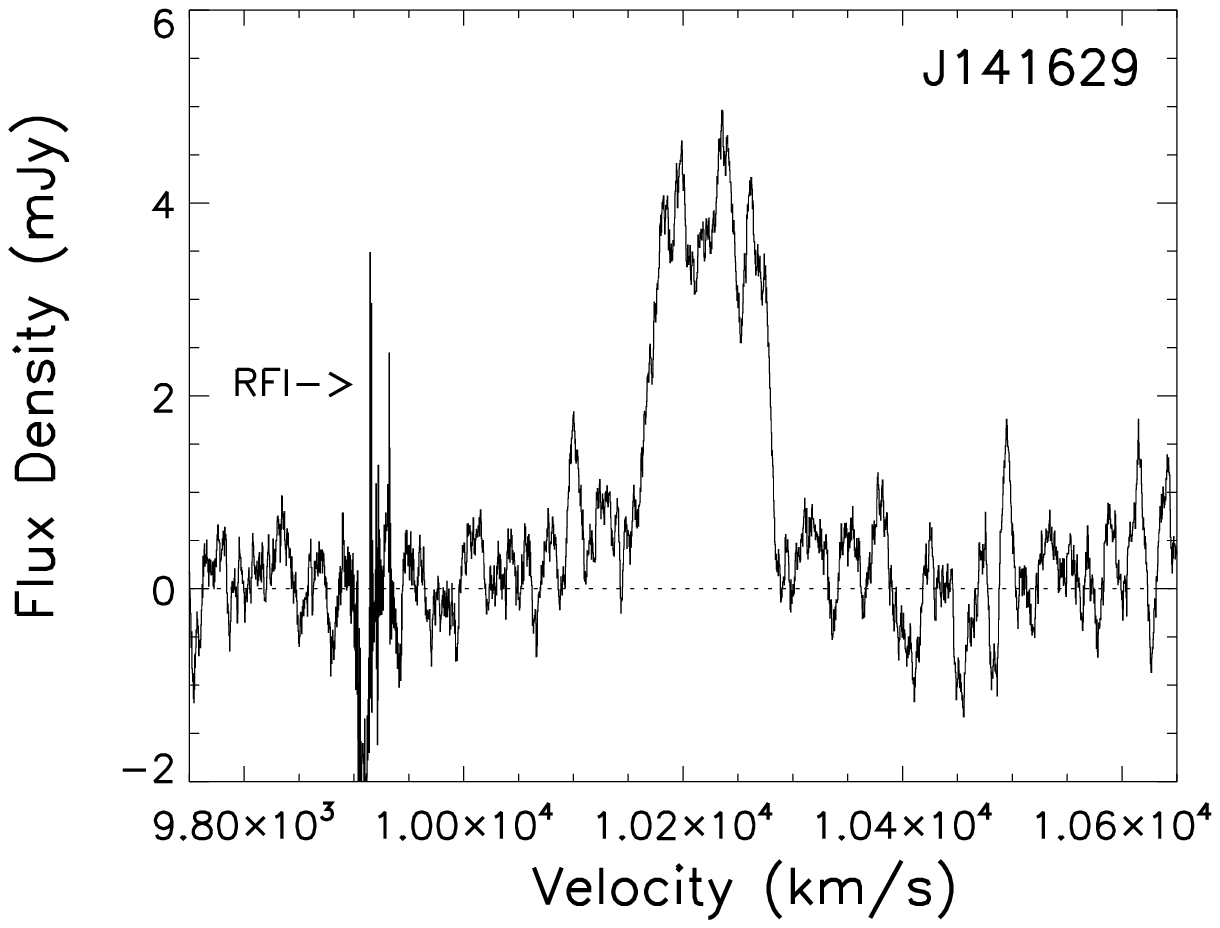} 
\includegraphics[trim= 0mm 0mm 0mm 0mm, clip=true, scale=0.46,angle=0]{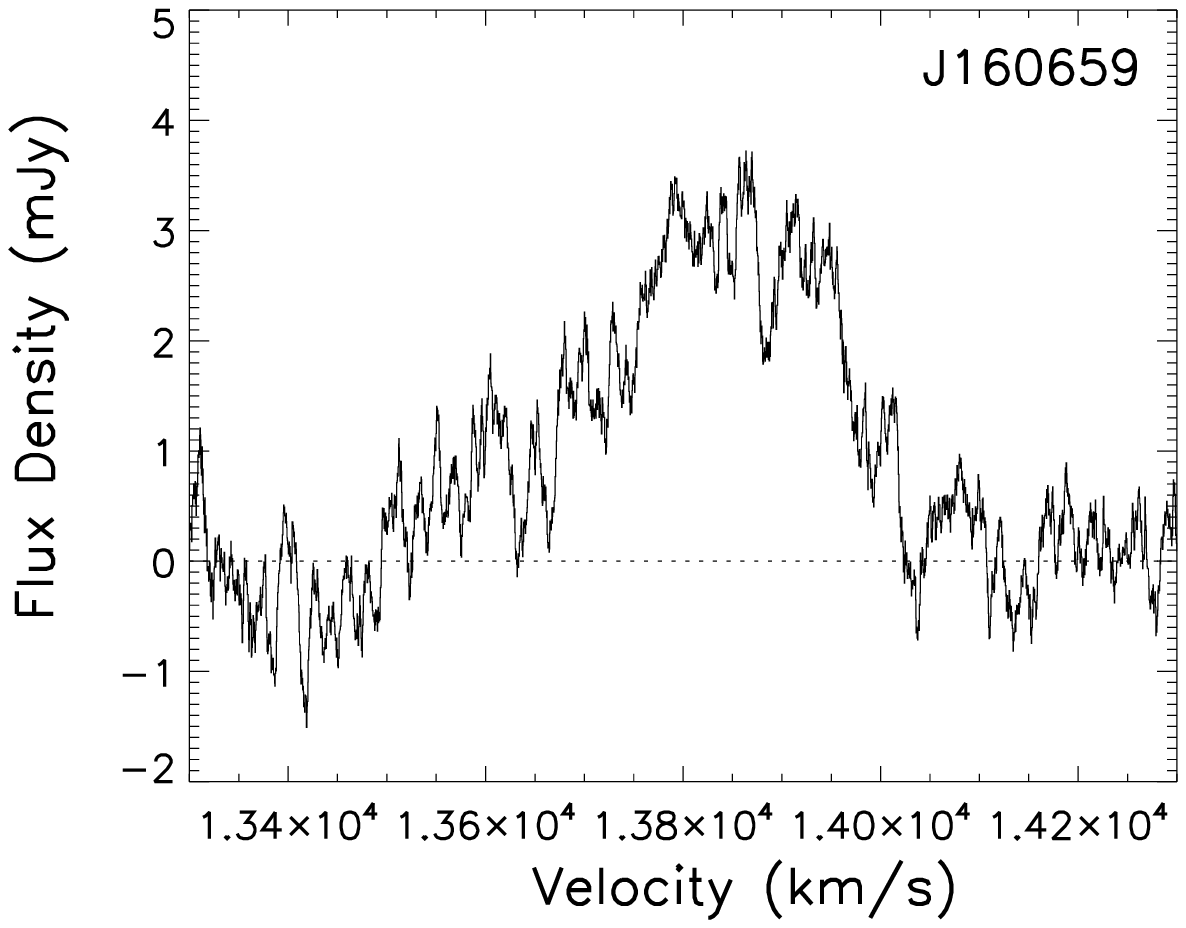} 
\caption{GBT \textsc{Hi} spectra towards background quasars probing three foreground galaxies from the survey by B11. The GBT beam covered the quasar sightline along with the foreground galaxy. The spectra show strong \textsc{Hi} emission features believed to be associated with the ISM of the foreground galaxy. 
The spectra towards quasar J122106+454852 probing galaxy UGC~7408 shows the data at original resolution. The spectra towards the other two sightlines probing galaxies J141629+372120 (labeled as J141629) and J160659+271642 (labeled as J160659) are smoothed by 30 pixels ($\equiv$ 10.6~\kms) to increase signal-to-noise ratio.
No clear indication of 21cm \textsc{Hi} absorption were seen in the raw or the smoothed data. The large GBT beam of FWHM 9.$^{\prime}$1 makes it impossible to interpret the physical significant of the non-detections of \textsc{Hi} absorption towards these quasar sightlines. One explanation could be that there is an absence of gas in the region probed by the sightlines. However, it is also possible that 21~cm absorption produced towards the quasar sightlines are being filled-in by \textsc{Hi} emission associated with gas elsewhere in the galaxies. Higher spatial resolution is warranted to distinguish between the two possible scenarios, thus our motivation for obtaining higher spatial resolution VLA observations. }
  \label{GBT_spectra}
\end{figure*}

 \begin{deluxetable*}{lcccc  rccccccrl}
\tablenum{1}
\tabletypesize{\scriptsize}
\tablecaption{Galaxies and Their Properties.  \label{tbl1-targets}}
\tablewidth{0pt}
\tablehead{
\colhead{Foreground Galaxy} & \colhead{Redshift} &\colhead{Stellar Mass} &\colhead{Metallicity\tablenotemark{a} }  & \colhead{\textsc{Hi}  Mass}  & \colhead{$v_{\rm HI peak}$\tablenotemark{b}} & \colhead{$\Delta v_{\rm HI}$\tablenotemark{c}} &\colhead{}  \\
\colhead{} &  \colhead{}  &  \colhead{(Log M$_{\odot}$)}&\colhead{(12+log(O/H))}  & \colhead{(Log M$_{\odot}$)}  & \colhead{($\rm km~s^{-1}$)} & \colhead{($\rm km~s^{-1}$)} &\colhead{} & \colhead{} }
\startdata
 J122115.22+454843.2\tablenotemark{d}  & 0.0015& 7.4 & 8.3\tablenotemark{e}  & 8.3 & 460 & 417 - 507 \\
 J141629.25+372120.4 & 0.0341& 9.3 &8.6 &9.4  &  10235 & 10070 - 10300\\
 J160659.13+271642.6 & 0.0462 & 9.3 & 8.4 &9.9 &  13860 &  13500 - 14030 \\
\enddata
\tablenotetext{a}{Metallicity based on N2 index \citep{pittini04}. The metallicity was measured from emission lines from the central region of the galaxy and may be considered as the upper limit for the gas probed by the quasar sightlines.}
\tablenotetext{b}{Velocity corresponding to the peak of the \textsc{Hi} profile.}
\tablenotetext{c}{Velocity width (full) of the emission profile.}
\tablenotetext{d}{Also known as UGC~7408.}
\tablenotetext{e}{Based on values published by \citet{kennicutt08}.}
\end{deluxetable*}

The accretion and subsequent condensation of inflowing ionized gas, with temperatures of $\rm 10^4$ to $\rm 10^6$~K, from the intergalactic medium (IGM) and circumgalactic medium (CGM) into the interstellar medium (ISM) of galaxies, with temperatures of 10$\rm ^{2}$ to 10$\rm ^{4}$~K, is a crucial aspect of the baryon cycle. 
While our theoretical understanding of gas inflow into galactic halos has progressed tremendously in the last couple of decades \citep[including work by][]{birn03, maller04, keres05, ford13}, the processes that enable the gas to actually condense and achieve physical conditions where stars can form are not yet well understood.
On the observational front, several studies have confirmed the presence of vast reservoirs of cool gas  \citep[$\rm 10^9~M_{\odot}$ of gas at $\rm 10^{4-5}$~K;][]{werk14} and metals \citep{wakker09, prochaska11, tumlinson11b, borthakur13,bordoloi14} in the CGM of galaxies of all types. 
Several studies have found  correlations between the kinematics of cooler and hotter CGM gas phases \citep{tripp08, tripp11, meiring13}.
Similarly, evidence of extended cool rotating disks has been seen using probes such as \ion{Mg}{2} absorbers  \citep{steidel02,kacprzak10}.

Where does the  infalling/circumgalactic gas condense into interstellar medium? How is that process regulated? And what do the properties of neutral gas in extended disks or extraplanar gas teach us about this process? 
In order to answer these questions and to understand the connection between the process of condensation and the properties of the ISM, detailed surveys of cold gas properties in the outer disks of galaxies are needed.
For this purpose, \HI\ 21~cm absorbers have been used by multiple teams to study properties of cold gas over a wide range of redshifts \citep[e.g.,][and references therein]{kanekar97,lane01,vermeulen03, darling04, keeney05, gupta07, gupta09, gupta10, gupta13, borthakur10b, borthakur11}.

\begin{deluxetable*}{lcccccccccccrl}
\tablenum{2}
\tabletypesize{\scriptsize}
\tablecaption{Targeted Galaxy-Quasar Pairs.  \label{tbl2-sightlines}} 
\tablewidth{0pt}
\tablehead{
\colhead{\#} &\colhead{B11\# \tablenotemark{a}} &\colhead{Foreground Galaxy} & \colhead{Redshift}  & \colhead{Background Radio Source} & \colhead{$S_{\rm FIRST}$} &\colhead{$\rho$} &\colhead{$ \Sigma_{SFR}$\tablenotemark{b}} \\
\colhead{} & \colhead{} & \colhead{} &  \colhead{}  &  \colhead{} & \colhead{(mJy)} &\colhead{(kpc)} & \colhead{$\rm (M_{\odot}~yr^{-1}~kpc^{-2})$} }
\startdata
1 & 12&J122115.22+454843.2\tablenotemark{c}  & 0.0015&  122105.480+454838.80 & 53.69 & 3.3  &  - \\
2 & 13&                   ''                                             &     ''      &   122106.854+454852.16 & 21.46 & 2.8 & $\rm <3.46 \times 10^{-4}$ \\
3 & 14&                    ''                                            &     ''      &   122107.811+454908.02 & 12.82 &2.6 &  - \\
4 & 21&J141629.25+372120.4                              & 0.0341&  141631.039+372203.01 & 30.71 & 32.4 & $\rm <2.34 \times 10^{-4}$ \\ 
5 & 22&                     ''                                           &            & 141630.672+372137.09 & 30.53  &16.2 &  - \\
6 & 23& J160659.13+271642.6                             & 0.0462& 160658.315+271705.86 & 141.01 & 23.3 &  - \tablenotemark{d}\\
\enddata
\tablenotetext{a}{Sightline I.D. from Borthakur et al. 2011} 
\tablenotetext{b}{SFR estimated from H$\alpha$ emission towards the optical QSO using SDSS3 spectra.}
\tablenotetext{c}{Also known as UGC~7408}
\tablenotetext{d}{Measurements could not be made due to the presence of sky lines at the wavelength of H$\alpha$.  \\}
\end{deluxetable*}

\begin{deluxetable*}{ccccccccccccrl}
\tablenum{4}
\tabletypesize{\scriptsize}
\tablecaption{HST Cosmic Origins Spectrogrph Follow-up Observations of UV-bright Background QSOs. \label{tbl3-hst_obs}}
\tablewidth{0pt}
\tablehead{
\colhead{Field} &\colhead{Program ID} &\colhead{Grating\tablenotemark{a} } &\colhead{Exposure}&\colhead{}  \\
\colhead{} & \colhead{} & \colhead{} &  \colhead{(s)}  }
\startdata
J104257.58+074850.5  & 12467 &  G140L & 2221\\
J122115.22+454843.2  & 12486 & G130M & 8085\\
\enddata
\tablenotetext{a}{ The far-ultraviolet (FUV) detector of the COS is sensitive to wavelengths between 900 and 2150~$\rm \AA$ and has medium-resolution (R $\sim$ 20,000) and low-resolution (R $\sim$ 3,000) gratings.}
\end{deluxetable*}

In the last five years two low-z surveys by \citet{gupta10} and \citet[][B11 hereafter]{borthakur11} attempted a census of cold gas detection rate as a function of impact parameter from the host galaxy.
These studies used radio-bright quasars to probe low-z galaxies at impact parameters ranging from 11-53~kpc and 2-100~kpc respectively. Due to rarity of small impact parameter quasar-galaxy pairs, the sample included a wide variety of galaxies with majority of them being sub-L$*$. Despite the inhomogeneity in their galaxy properties, the two studies have provided independent estimation of covering fraction of cold gas to be $\approx$50\% within $\sim$20~kpc. The detection of cold gas beyond that distance is exceedingly rare. However, it is worth noting that these surveys sparsely sampled the impact parameter range and did not take into account the effects of galaxy orientation. On the other hand, the number density and cross-section of Damped Lyman Alpha (DLA; i.e. absorbers with N(HI)$\rm\ge ~2\times10^{20}~cm^{-2}$) systems suggests a covering fraction of \HI to be around 50\% within 20~kpc of a galaxy \citep[see Section 4.1 in B11; Section 6.1 in ][]{schaye07}. 
Interestingly, the Milky Way's atomic gas disk with \HI column densities to qualify as a DLA is about 20~kpc \citep[][and references therein]{strasser07,kalberla09}, although the disk extends to about twice that distance at much lower column densities \citep[see Figure 8][]{kalberla09}.

In this paper, we present follow­-up observations of three of the galaxies from our previous \HI absorption survey (B11). In our previous work, we reported nondetection of \HI absorption in two of the targets based on observations with the Green­ Bank Telescope (GBT). However, we noted that due to the large GBT beam, it is possible that \HI absorption against the background source is filled-in by emission from elsewhere in the galaxy. The GBT spectra obtained by B11 are shown in Figure~1.  These single dish data were obtained using ON-OFF position switching mode. Therefore, in order to confirm the absence of absorption, we conducted follow-up observations with the Karl G. Jansky Very Large Array (VLA) of the NRAO\footnote{The National Radio Astronomy Observatory is a facility of the National Science Foundation operated under cooperative agreement by Associated Universities, Inc.} in B-configuration. These new observations provide almost two orders of magnitude higher spatial resolution than the GBT. In addition, we also obtained Hubble Space Telescope (HST) ultraviolet (UV) spectra of the background quasi-stellar objects (QSOs) in order to get an independent measurement of \HI column density.

We proceed by describing our targets, followed by details of the observations and data analysis in Section 2. The results and a discussion on the implications of the findings are presented in Section 3. Finally, we conclude in Section 4 and comment on the potential of similar studies with upcoming facilities in the future. 
The cosmological parameters used in this study are $H_0 =70~{\rm km~s}^{-1}~{\rm Mpc}^{-1}$, $\Omega_m = 0.3$, and $\Omega_{\Lambda} = 0.7$.

\begin{deluxetable*}{lcccccccccccrl}
\tablenum{3}
\tabletypesize{\scriptsize}
\tablecaption{Description of VLA Observations and the Data Products.  \label{tbl2-vla_obs}}
\tablewidth{0pt}
\tablehead{
\colhead{Field} &\colhead{Duration} &\colhead{Bandwidth} & \colhead{Bandwidth} &\colhead{Channel Width}& \colhead{Beam Size\tablenotemark{a}} & \colhead{$\sigma_{channel}$} & \colhead{$\sigma_{cont. map}$\tablenotemark{b} }   \\
\colhead{} & \colhead{(hrs)} & \colhead{(MHz)} &  \colhead{($\rm km~s^{-1}$)}  &  \colhead{($\rm km~s^{-1}$)} & \colhead{(arcsec$\times$arcsec)} & \colhead{(mJy/beam)}  & \colhead{(mJy/beam)} }
\startdata
J122115.22+454843.2\tablenotemark{c}  & 2.0  & 1.0 & 211 & 0.8  & 7.45 $\times$ 5.58 & 3.06  &  0.292 \\
J141629.25+372120.4 & 8.0 & 2.0 & 436 & 1.7 &  7.13 $\times$ 5.22 & 1.0  & 0.078    \\
J160659.13+271642.6 & 8.0 & 4.0 & 883 & 3.4 & 8.35 $\times$ 5.94 & 0.9  & 0.102  \\
\enddata
\tablenotetext{a}{Reported for the line image. The variation in beam size between continuum and the line image is less than 10\%.}
\tablenotetext{b}{Average of all channel.}
\tablenotetext{c}{Also known as UGC~7408.}
\end{deluxetable*}

\section{OBSERVATION \label{Sec:uvobservations}}

\subsection{Targets \label{Sec:VLA_targets}}

\subsubsection{Targets for High Spatial Resolution 21~cm HI Imaging with the VLA \label{Sec:VLA_targets}}

We carried out follow up VLA observations of three galaxies with radio bright background sources.  Properties of the target galaxies such as their redshifts, stellar and \HI masses, central metalicities, \HI velocity centroids and widths are provided in Table~\ref{tbl1-targets}.
The GBT spectrum of the three galaxies obtained by B11 shows strong \HI emission. The measured \HI masses indicate that these galaxies have more \HI than stars. Based on the \HI mass to disk size relationship derived by \citet{swaters02}, these galaxies are expected to have extended \HI\ disks well beyond their optical disks.

These three galaxies are probed by six sightlines corresponding to six radio bright background sources. Information on each of the sightlines including their IDs from the study by B11, position of the background radio sources, their 20~cm fluxes, and impact parameters are provided in Table~2.  
Three of the sightlines are QSOs with optical spectroscopic data from the Sloan Digital Sky Survey (SDSS).
The non-detection of any H$\alpha$ in emission at the redshift of the foreground galaxy in the QSO spectra provides us with an upper limit on the star formation rate surface density ($\Sigma_{\rm SFR}$).

Since all the galaxies were found to have strong 21~cm \HI emission in the GBT spectra, it is impossible to rule out absorption being filled-in by emission from a different spatial location. Hence, we carried out high spatial resolution VLA observations to confirm the non-detections at the location of the background radio sources. The VLA in B-configuration is most appropriate for such a study as the size of the synthesized beam ($\sim\rm 8^{\prime\prime}\times 6^{\prime\prime}$) matches the extent of the background continuum sources and is hence most conducive to detecting absorption features. For example, the spatial resolution achieved in our imaging (column 6 of Table~3
) is better than that prescribed by \citet[][resolution of $<10^{\prime\prime}$]{dickey00} for extragalactic sources.

\subsubsection{Targets for Follow-up  Ultraviolet Spectroscopy with the HST \label{Sec:HST_targets}}

We obtained UV spectra for two of the QSO sightlines from our original sample that had confirmed 21~cm absorbers. The purpose of these observations was to detect and measure the \Lya ($\lambda \rm 1215.670~\AA$) transition associated with the 21~cm absorbers and hence provide a measure of the \HI column density independent of any assumption regarding the spin-temperature of the gas. The combination of \HI absorption properties from the 21~cm hyperfine transition and the \Lya transition can be used to measure the column density and spin temperature of the absorber simultaneously. The QSOs were observed with the Cosmic Origins Spectrograph (COS) aboard the HST under programs 12467 and 12486. One of the sigtlines, SDSS~J122115.22+454843.2, is the optical/UV counterpart of the radio sightline \#2 (see Table~\ref{tbl1-targets}). The other sightline is towards  SDSS~J104257.58+074850.5 probing galaxy GQ1042+0747 at an impact parameter of 1.7~kpc \citep{borthakur10b}.
The setups for the HST observations are presented in Table~4. 

\begin{figure*}
\figurenum{2}
\includegraphics[trim= 10mm 40mm 20mm 20mm, clip=true, scale=0.45,angle=90]{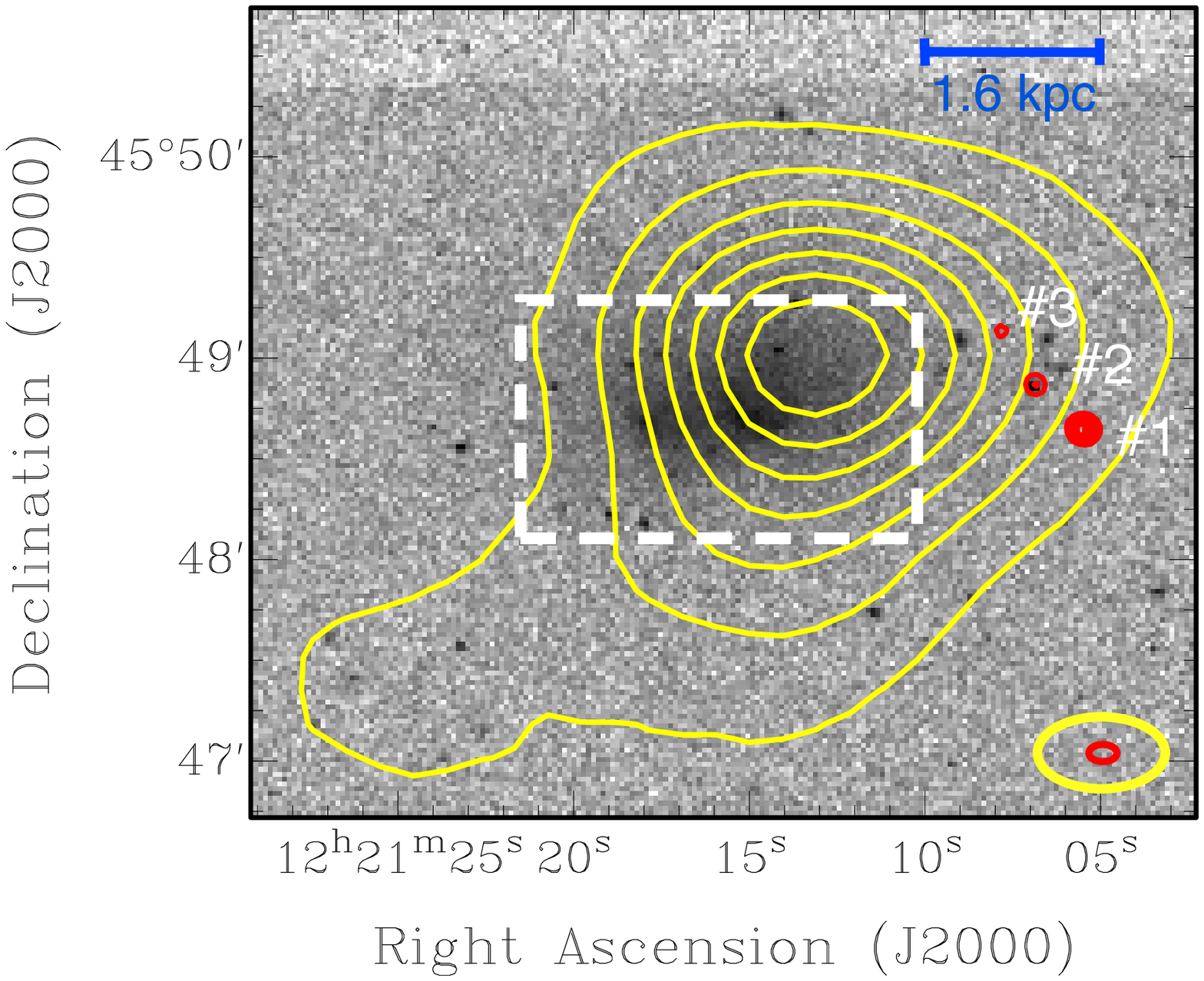} 
\includegraphics[trim=13mm 0mm 13mm 0mm, clip=true, scale=0.37,angle=-0]{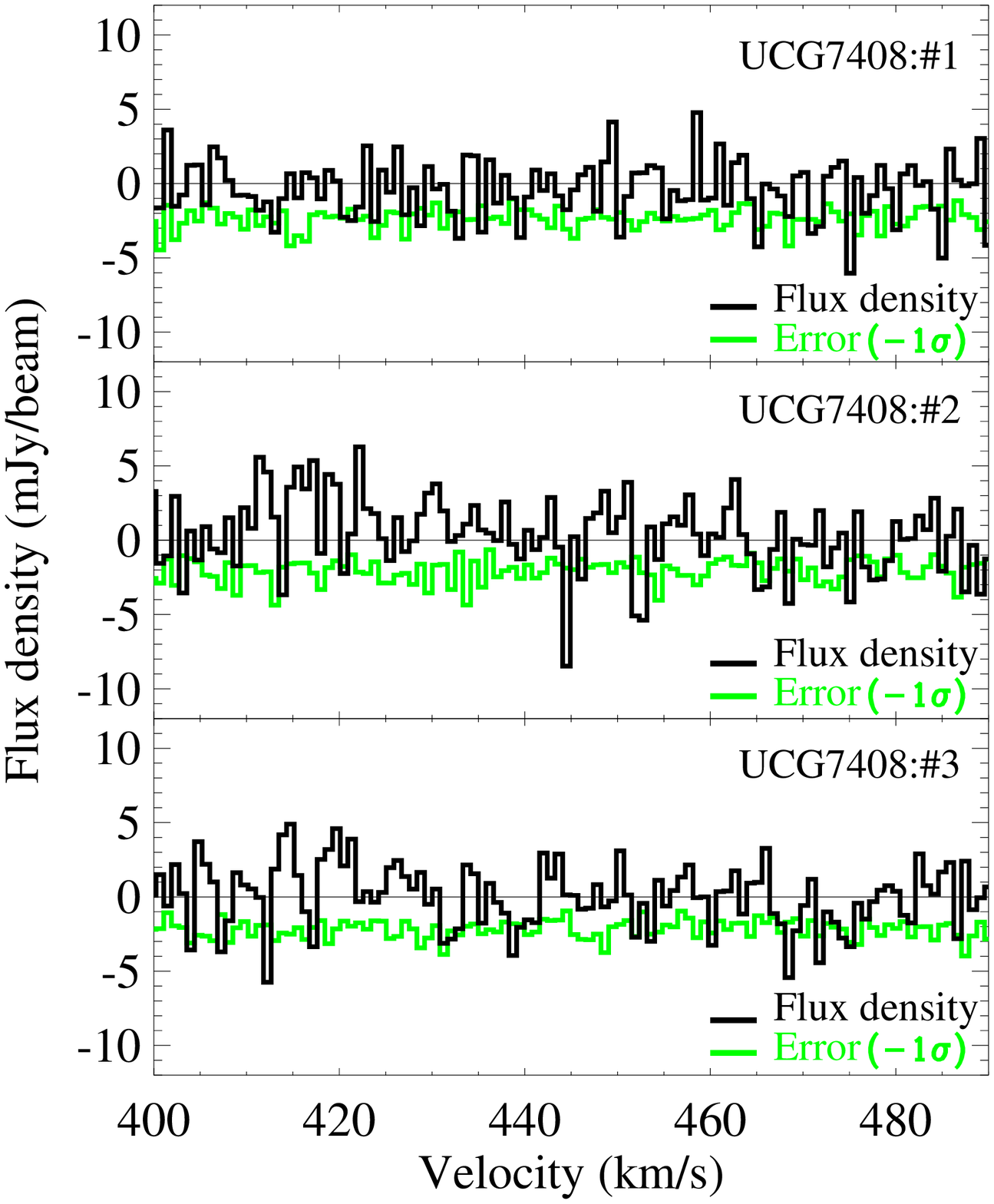} 
\caption{Left: SDSS r-band image of UCG~7408 in grayscale with VLA B-configuration continuum map and VLA D-configuration \textsc{Hi}  map (B11) shown as contours in red and yellow, respectively.  The target galaxy, UCG~7408, is marked with a dashed white rectangle and the sightlines are labeled. The red contours show VLA B-configuration continuum map with flux levels of 10, 20, 30, 40, and 50 mJy/beam. The VLA D-configuration \textsc{Hi}  21~cm column density are shown in yellows at levels corresponding to   7.2, 14.4, 21.6, 28.8, 36.0, 43.2, and 50.4 $\rm \times~10^{19})~ cm^{-2}$, respectively. The synthesized beam for the D-configuration map (67$^{\prime\prime}$.51 $\times$ 45$^{\prime\prime}$.31) is shown in yellow and that of the B-configuration is shown in red at the bottom right corner. 
Right: Spectra extracted at the position of the background sources (sightlines \# 1, 2, and 3) are presented on the right panel. The errors in the measurements corresponding to standard deviation in flux density for each of the channel maps, derived using the imaging package in CASA, are shown in green. 
The errors are plotted in negative units (i.e. $-1\sigma$) for easy in visually estimating the strength of absorption features. 
A single pixel feature was detected in the spectrum towards sightline 2 at a velocity of 444.5~\kms. The spatial map of the \textsc{Hi}  in absorption is presented in the Figure~\ref{ucg7408_zoom}. We do not see any absorption feature at strength 3$\sigma$ of higher towards he other two sightlines probing the \HI disk of UGC~7408.}
\label{TarA}
\end{figure*}

\begin{figure*}
\figurenum{3}
\includegraphics[trim=20mm 10mm   20mm 60mm, clip=true, scale=0.42,angle=90]{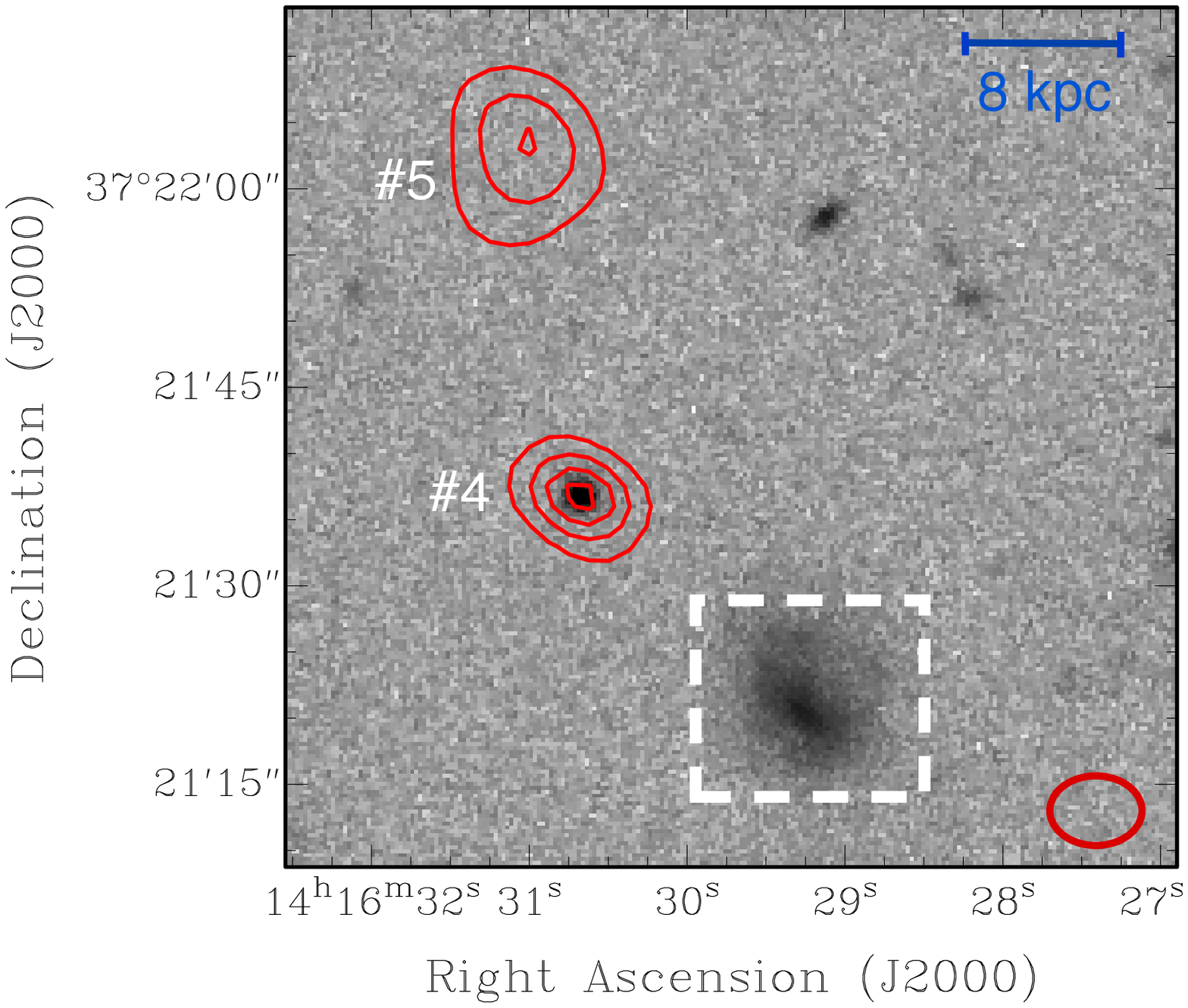} 
\includegraphics[trim=15mm 0mm 13mm 0mm, clip=true, scale=0.37,angle=-0]{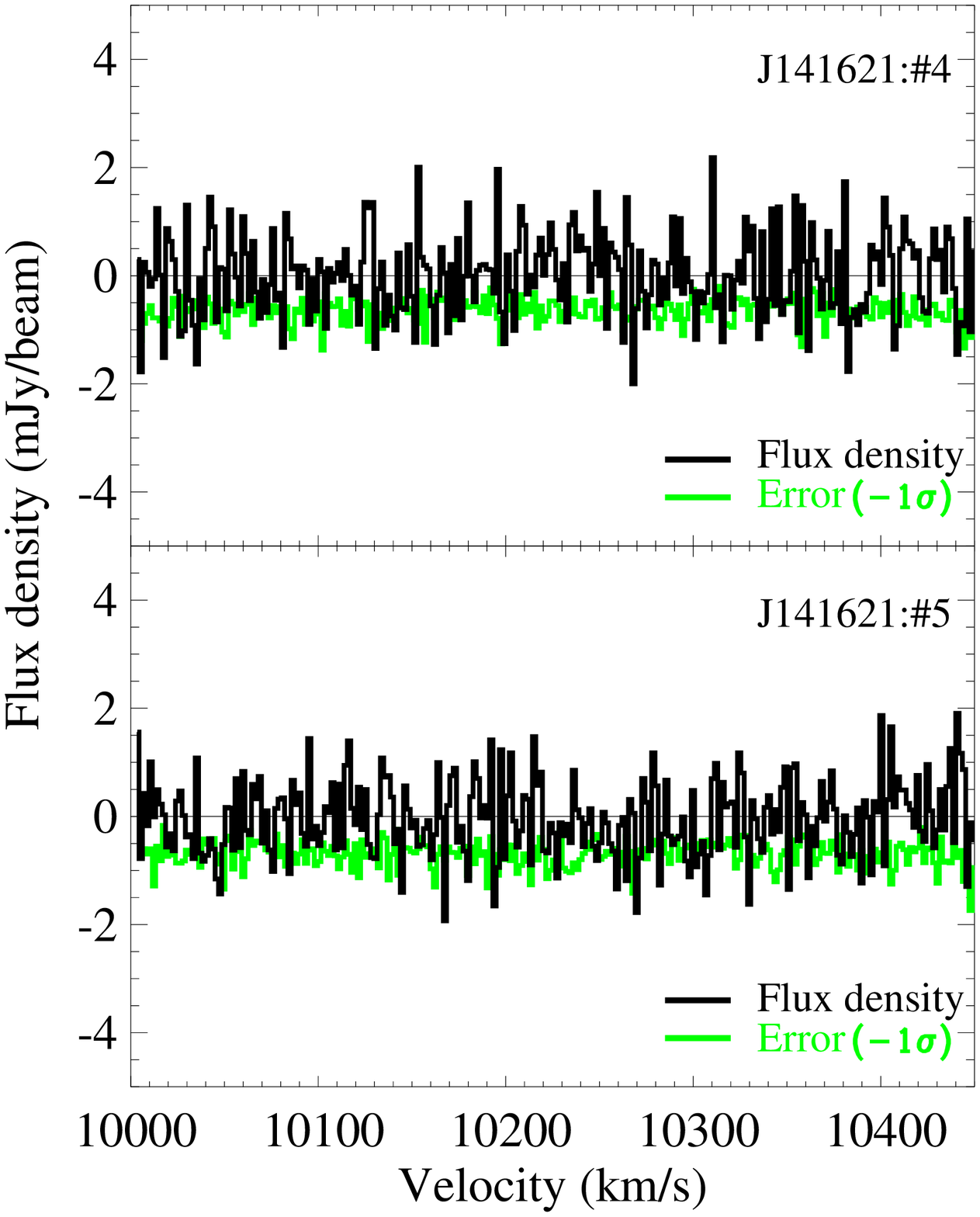} 
\caption{Left: SDSS r-band image of SDSS~J141629.25+372120.4 in grayscale overlaid with contours of VLA B-configuration continuum map shown in red.  The target galaxy, J141629, is marked with a dashed rectangle in white and the sightlines are labeled. The red contours are at flux levels of 5, 10, 15 and 20 mJy/beam. The beam is shown at the bottom right corner in red and the physical scale at the restframe of the target galaxy is shown on the top right-hand corner. Right: Spectra extracted at the position of the background sources (sightlines \# 4 and 5) are presented in the right panel. The standard deviation in flux density ($-1\sigma$) in each of the channel maps are plotted in green. We do not see any absorption feature at strength 3$\sigma$ or higher for both the sightlines. }
\label{TarB}
\end{figure*}

Unfortunately, the COS spectra for both the targets had insufficient flux at the rest-frame wavelength of \Lya for the targeted galaxies. As a result no measurements could be made. The drop in UV flux of the QSOs is consistent with the presence of Lyman Limit Systems (LLS) at higher redshift towards these sightlines that have consumed most of the flux at shorter wavelengths. 
Therefore, for the remainder of the paper, we concentrate on the analysis and results from the 21~cm follow-up observations.

\subsection{Data Acquisition and Analysis \label{sec:data_analysis}}

\subsubsection{Very Large Array Observations  \label{sec:obs_vla}}

We carried out a total of 18 hours of observations with the VLA in B-configuration under program 10C-120. We observed the field towards SDSS~J122115.22+454843.2 (UCG~7408 hereafter) for 2~hours and the fields towards SDSS~J141629.25+372120.4 (J141629 hereafter) and SDSS~J160659.13+271642.6 (J160659 hereafter) for 8~hours each. The observations were carried out in dual polarization mode with a total bandwidth of 1, 2, and 4 Megahertz (MHz), respectively. This setup was chosen to yield velocity coverage of 211, 436, and 883 \kms respectively.  
The VLA correlator, WIDAR, was set up to deliver 256 spectral channels for each of the used bandwidths, resulting in a velocity width of 0.8, 1.7, and 3.5 \kms per channel, respectively. 
WIDAR being an XF correlator, the frequency resolution of the raw (unbinned and unsmoothed) spectra is 1.2 times the channel width \citep{rots82}.

The data were reduced and calibrated following the standard VLA calibration and imaging procedures using the Common Astronomy Software Applications package (CASA). The data were imaged using a 
``natural" weighting scheme for spectral line images. The natural weighting was preferred as it offered the highest possible signal to noise ratio in the images. Details of the observations and data products are provided in Table~3.

\subsubsection{ Existing Very Large Array Data  }\label{sec:obs_vla_D}

 We also present existing VLA D-configuration data that were obtained under legacy I.D. AY190 and were first published by B11. 
The data were obtained in 2008 in dual polarization mode with a bandwidth of 1.5 MHz and 128 spectral channels using the old VLA correlator. These correspond to a velocity coverage of 330~\kms and a channel width of 2.6~\kms, respectively.
The data were reduced and calibrated following the standard calibration and imaging procedures using NRAO's Astronomical Image Processing Software (AIPS). The spatial resolution achieved with the uniform weighting of the data is 52.$^{\prime\prime}$17$~\times~$38.$^{\prime\prime}$11, which is not enough to resolve the background radio source into three individual sources. And hence, detailed spatial information on the associated \HI absorption could not be made with these data, thus requiring the need for higher spatial and spectral resolution B-configuration data.

\section{RESULTS AND DISCUSSION \label{Sec:discussion}}

\begin{figure*}[!h]
\figurenum{4}
\includegraphics[trim=30mm 40mm   30mm 40mm, clip=true, scale=0.49,angle=90]{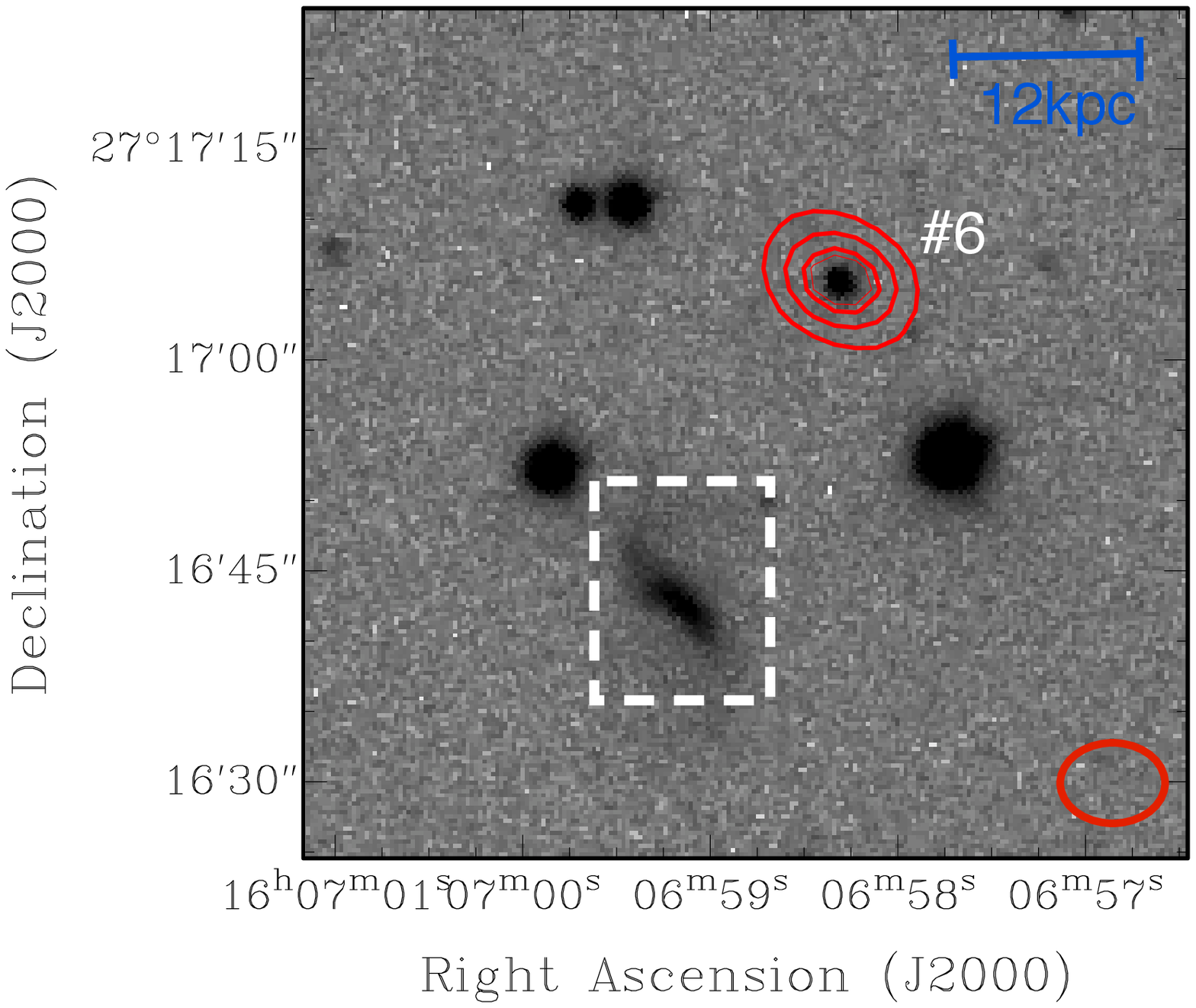} 
\includegraphics[trim=15mm 0mm 00mm 0mm, clip=true, scale=0.45,angle=-0]{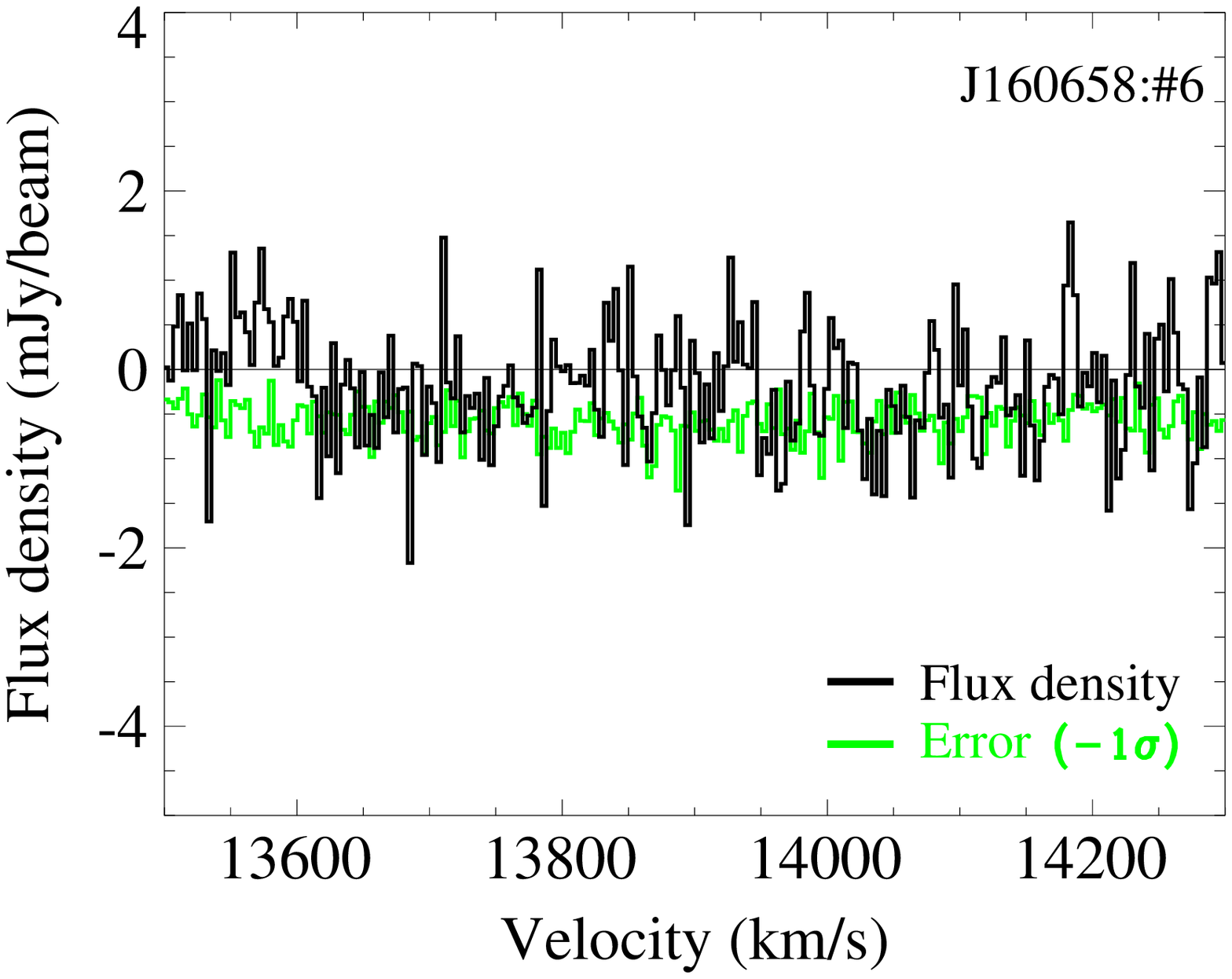} 
\caption{Left: SDSS r-band image of SDSS~J160659.13+271642.6 in grayscale overlaid with contours of VLA B-configuration continuum map shown in red.  The target galaxy, J160659, is marked with a dashed rectangle in white and the sightlines are labeled. The red contours are at flux levels of 50, 100, 150 and 170 mJy/beam. The VLA beam is shown at the bottom right corner in red and the physical scale at the rest-frame of the target galaxy is shown on the top right-hand corner. Right: Spectrum extract at the position of the background sources (sightline \# 6) is presented on the right panel.  The standard deviation in flux density ($-1\sigma$) in the channel maps is plotted in green. We do not see any absorption feature at strengths 3$\sigma$ of higher towards this sightline. This being the background source with the largest integrated  flux among the six sightlines discussed in this paper, we estimate the most stringent optical depth limit of  $\tau_{\#6}\le 0.03$ for this target. }
\label{TarC}
\end{figure*}

In this section we present the continuum images and \HI spectra from our VLA observations. Figures~\ref{TarA}-\ref{TarC} show 21~cm radio continuum images overlaid as red contours on the SDSS r-band images in greyscale. Figure~\ref{TarA} also shows the VLA D-configuration \HI emission map overlaid as yellow  contours (first published by B11). The flux density measured for each of the background radio sources (continuum) is provided in column~4 of Table~\ref{tbl4-results}. Spectra extracted from each of the sightlines are shown in black on the right panels. The standard deviation in flux density per channel is plotted green. The errors are plotted in negative units (i.e. $-1\sigma$) for easy in visual identification of significant absorption features.

 \begin{deluxetable*}{llccccccccccrl}
\tablenum{5}
\tabletypesize{\scriptsize}
\tablecaption{Optical Depth and Column Density Measurements. \label{tbl4-results}}
\tablewidth{0pt}
\tablehead{
\colhead{\#} & \colhead{Background Radio Source}&\colhead{$\rho$}   & \colhead{$S_{\rm 1.4~GHz}$} & \colhead{$\sigma_{4km/s}$} & \colhead{Abs strength} & \colhead{$\tau_{3\sigma}$} & \colhead{N(HI)$_{3\sigma}$\tablenotemark{a} } \\
\colhead{} & \colhead{} & \colhead{(kpc)} & \colhead{(mJy)} &\colhead{mJy}  &  \colhead{(mJy)} & \colhead{} &  \colhead{($\rm \times 10^{19}cm^{-2}$)} }
\startdata
1 & 122105.480+454838.80 & 3.3   & 65.18 &  1.42 & $<$4.26 & $<$0.07 & $<$ 2.55\\
2 & 122106.854+454852.16 & 2.8   & 24.72 &  1.42 & - & - & - \\
2\tablenotemark{b}  & 122106.854+454852.16\tablenotemark{b} & 2.8   & 13.98 &  - & 10.41\tablenotemark{c} & 1.39\tablenotemark{c}  & ~~6.47 \\
3 &122107.811+454908.02  & 2.6   & 13.44 &  1.42 & $<$4.26& $<$0.39 & $<$14.22\\

4 & 141630.672+372137.09 & 16.2 & 32.79 & 0.65  & $<$1.96& $<$0.06 & $<$2.19\\
5 & 141631.039+372203.01 & 32.4 &25.30  & 0.65  & $<$1.96& $<$0.08 & $<$2.92\\

6 & 160658.315+271705.86 & 23.3 & 190.17 & 0.83 & $<$2.49& $<$0.01& $<$0.36\\
\enddata
\tablenotetext{a}{Assuming a line width of 4~\kms , a spin temperature $T_s=$50~K, and a covering fraction, $f=1$.}
\tablenotetext{b}{Region where the absorption was detected (see Figure~2 for details).}
\tablenotetext{c}{At the intrinsic resolution of the data (channel width of 0.84~\kms). }
\end{deluxetable*}

  \subsection{Cold Gas in UGC~7408}

We detected a narrow absorption feature associated with UGC~7408 in our VLA B-configuration spectra towards sightline \#2. The absorption feature peaks at a velocity of 444.5~\kms and was measured to have a full-width at half maximum (FWHM) of 1.1~\kms. The width of the feature is comparable with our spectral resolution. No significant absorption features were seen towards sightlines \#1 and  \#3.

\begin{figure}
\figurenum{5}
\center
\includegraphics[trim=0mm 0mm 0mm 5mm, clip=true, scale=.35]{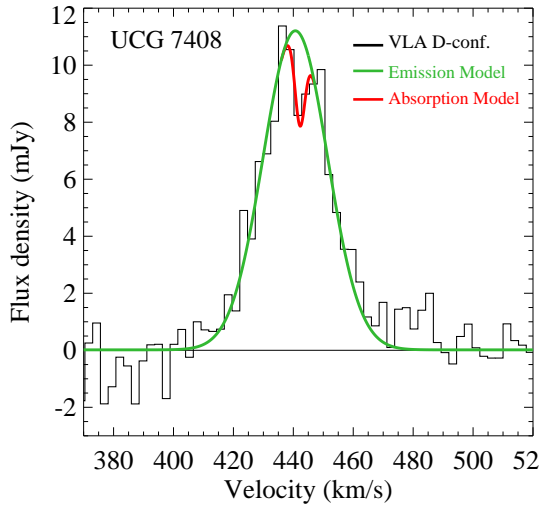} 
\caption{VLA D-configuration \textsc{Hi} spectrum extracted from a region the size of the synthesized beam centered at the quasar (first published by B11). The data show a Gaussian \textsc{Hi} emission feature with an absorption feature (FWHM=4.75 \kms ) superimposed at the peak of emission. B11 concluded that the broad Gaussian emission feature was produced by the atomic gas that filled-in the beam, whereas the gas towards the quasar sightline, covering only a small fraction of the beam, produced the superimposed absorption. }
\label{ucg7408_D}
\end{figure}

Previously, B11 had found an \HI absorption feature associated with UGC~7408 towards the background quasar encompassing sightlines \# 1, 2, and 3. The absorption was found as dip on an otherwise Gaussian emission profile. The VLA D-configuration spectrum extracted from a region the size of the synthesized beam centered at the quasar is shown in Figure~\ref{ucg7408_D}.
 B11 concluded that the atomic gas that filled-in the beam contributed to the broad Gaussian emission feature, whereas the gas towards the quasar sightline, covering a small  fraction of the D-configuration beam, produced the superimposed absorption. By subtracting the emission, as modeled by a Gaussian profile, these authors found the absorption feature to have a FWHM of 4.75~\kms at the spectral resolution of 3.1~\kms corresponding to a channel width 2.6~\kms \citep{rots82}. The feature peaked at  442.2~\kms. It is worth noting that the velocity centroid of the absorption feature is likely to be effected by errors in modeling the intrinsic emission. For example, any deviation from a Gaussian emission profile would introduce a spurious offset.

Our 21cm absorber towards sightline \#2 is consistent with the superimposed absorption feature detected by B11. The offset in the velocity centroids between the two detections is consistent with the difference in the spectral resolution between the two data sets (a factor of $\sim$2.5).  
This is also supported by the fact that the two absorption features have similar strengths that differ by less than 8\%. 
The \HI absorber detected in the D-configuration is broader and shallower than that detected in B-configuration.
Therefore, the D-configuration data are consistent with lower spectral resolution observations of an intrinsically narrow and deep absorption feature such as the one detected in B-configuration. While the two detections are consistent, the narrowness of the absorber clearly demonstrates the need for higher spectral resolution to obtain precise constraints on the physical conditions and kinematics of 21~cm absorbers.

\begin{figure*}
\figurenum{6}
\includegraphics[trim=30mm 104mm 30mm 100mm, clip=true, scale=1.15]{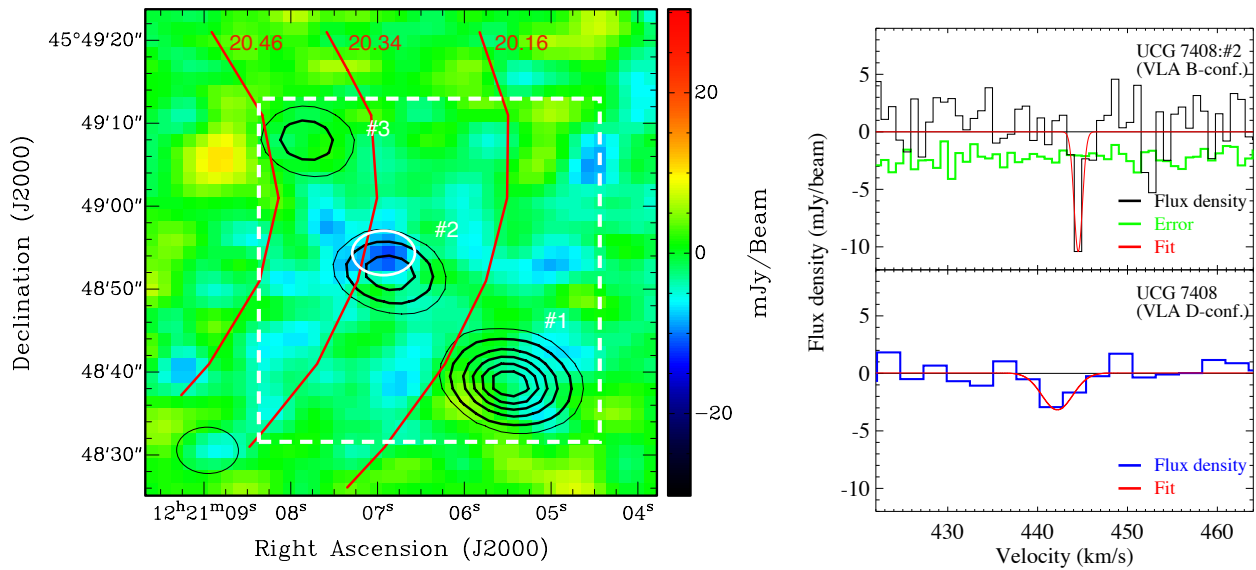} 
\caption{Left: 21~cm \textsc{Hi} absorption map of UCG~7408 corresponding to  the channel at velocity, $v=444.5$~\kms. The colors show the observed flux densities such that blue represents pixels with negative flux density i.e. {\color{blue}absorption},  yellow/red  represents pixels with positive flux density i.e. {\color{red} emission}, and green indicate pixels within the observed {\color{green} noise ($\pm 1\sigma$)} in the data. The background source is shown in black contours at flux density levels of 5, 10, 20, 30, 40, and 50 mJy/beam.  It is worth noting that we are sensitive to absorption only against the background radio source. However, we are sensitive to emission in the entire region. The synthesized beam size ($\equiv$230~pc$\times$173~pc in physical units) is shown in the lower left-hand corner. The red contours show the VLA D-configuration \textsc{Hi} emission map with contour levels indicating 14.4, 21.7, and 28.9 $\rm \times 10^{19}~cm^{-2}$. Spectra extracted from the regions marked in white are shown in the right panel.  Right: VLA B-configuration \textsc{Hi} spectrum extracted from the region marked in the white oval, the size of the beam, is shown in the top panel in black. The absorption has a peak depth of 10.41~mJy and the flux density of the background source at the same region is 13.98~mJy. The absorption feature is unresolved and the limiting FWHM of  was measured to be 1.1~\kms. This corresponds to a kinetic temperature, $\rm T_k\le$26~K. The VLA-D configuration \textsc{Hi} spectrum extracted from the white dashed rectangular region (similar in size to the D-configuration beam) is shown in the lower panel in blue. The feature has a FWHM of 4.75~\kms and a centroid at 442.2~\kms. The absorption spectrum was obtained by subtracting out the Gaussian emission profile from the raw spectrum as discussed in detail by B11.  }
\label{ucg7408_zoom}
\end{figure*}

Similar narrow \HI features have been detected in absorption in the Large Magellanic Cloud (LMC) \citep[1.4~$\rm km~s^{-1}$ by][]{dickey94}, and Small Magellanic Cloud (SMC) \citep[1.2~$\rm km~s^{-1}$ by][]{dickey00}.  
However, until recently most of the extragalactic \HI absorption surveys, owing to their low spectral resolution,  detected only absorbers with large widths. 
Therefore, it is possible that these studies might have missed absorbers such as this.

  \subsubsection{Covering Fraction of Cold Gas in UGC~7408} 
  
We present the \HI map from the channel corresponding to the velocity of the absorption feature (444.5~\kms) seen towards sightline \#2 in Figure~\ref{ucg7408_zoom}.
The background radio source is shown in black contours. 
The map is color coded to show absorption in blue, noise ($\rm \pm1\sigma$) in green, and emission in yellow/red. However, unlike emission, \HI absorption can only be measured in the region covered by the background radio source. 

Absorption associated with sightline \#2 can be seen in blue in Figure~\ref{ucg7408_zoom}, 
covering the Northeast quadrant of the background source. The physical size (area)  of the \HI cloud responsible for the absorption is $\sim \rm  0.04~kpc^2$. The implied covering fraction, $f$, of the background source by the cloud is $ \approx 25-30\% $.

\subsubsection{Physical Properties of the \HI Absorber}

A spectrum extracted from the position of the absorber (identified as a white oval of same size as the beam) is shown on the right. 
We measured the peak depth in flux density for the feature to be 10.41~mJy. 
For comparison, we also present in Figure~\ref{ucg7408_zoom} the D-­configuration absorption spectrum that was obtained by subtracting the \HI emission model from the spectrum in Figure~\ref{ucg7408_D}. The region of the galaxy from which the D­-configuration spectrum was extracted is shown as the white dashed rectangle on the image in the left panel of Figure~\ref{ucg7408_zoom}.
As noted earlier, the absorption feature seen in the B-configuration data is deeper and narrower than that seen in the VLA D-configuration data. 
The B­-configuration data have higher spectral and spatial resolution, however they have lower signal to noise ratio than the D­-configuration data. Nevertheless, the feature was confirmed at $> \rm 5\sigma$ at that channel or at $> \rm 3\sigma_{average}$, the average noise over all channel.


From our B-configuration data, we estimated the peak optical depth of this absorber to 1.37 using the following relation
\begin{equation}{\label{eq-tau}}
\tau= -~ln\Big(\frac{I_o-I_{abs}}{I_o}\Big)
\end{equation}
where $I_o$ is the flux density of the background quasar at the position of the absorber of 13.98 mJy (see Figure~\ref{ucg7408_zoom} left panel) and $I_{abs}$ is the peak strength of the absorption feature of 10.41~mJy. If we assume the width of the line is due to collisional excitation of the hydrogen atoms, we can relate the kinetic temperature of the gas to the width of the line as 
 \begin{equation}{\label{eq-kin_temp}}
 T_{k}\le 21.855~(\Delta ~v)^{2}
 \end{equation}
where $\Delta v$ is the FWHM of the line in \kms. In this case, $\Delta v =$1.1~\kms and implies a kinetic temperature, $ T_{k} \le 26.4$~K. This indicates that the temperature of this absorber is similar to that observed in cold neutral medium (CNM) of the LMC and SMC \citep{dickey94,dickey00}.

Assuming the kinetic temperature as a proxy for spin temperature (i.e. $ T_k=T_s=\rm 26.4~K$), we estimate the column density of \HI\ in this absorber to be $\rm 6.3\times10^{19}~cm^{-2}$ using the following expression
\begin{equation}{\label{eq-col_den}}
N({\rm H~I}) = 1.823 \times 10^{18} \frac{T_{s}}{f}\int \tau(v) dv ~ cm^{-2},
\end{equation}
where $\tau(v)$ is the 21~cm optical depth as a function of velocity in km~s$^{-1}$. From the Gaussian fit to the absorber, we estimate $\int\tau(v)~dv~= 1.3$. In this case, the covering fraction is unity, $f=\rm1$, as we are measuring optical depth, $\tau(v)$, at the region of the background source where the absorber was detected.

The column density of cold gas as derived from the absorption feature is significantly lower than the total \HI column density seen in emission.
By comparing the two, we find that the ratio of cold-to-total \HI column density associated with the absorber to be $\approx 30\%$. 
Folding in the covering fraction, we find that only $\approx10\%$ of the total \HI by mass exists in the cold phase.
The non-detections of \HI\ in absorption towards sightlines 1 and 3 provide upper limits on the column densities of 2.9 and 14.2~$\rm \times 10^{19}~cm^{-2}$ respectively. The same exercise of comparing these column densities to that observed in emission implies that the cold-to-total \HI\ towards sightlines 1 and 3 are $\lessapprox20\%$ and  $\lessapprox50\%$, respectively.

\subsubsection{Nature of \HI in UGC~7408} 

Our results suggest that the process of condensation is suppressed in UGC~7408. 
We found that a large fraction of atomic gas in the extended disk of this galaxy is warm with temperature much larger than a few 100~K.  
Therefore, we conclude that most of the atomic gas in the extended disk of UCG~7408 failed to condense into the atomic gas clouds of 100-50~K (typically found in the Milky Way ISM).
We did detect a small fraction of gas in the cold phase with temperature $\sim 25$~K. 
This is much lower than the temperatures commonly seen in the CNM of the Milky Way \citep[][]{heiles03, strasser07}.  In fact, at such low temperatures most of the gas in the Milky Way is in the molecular phase.
The existence of such cold atomic gas suggests that the fraction of \HI that was able to condense into the cold phase remained in the atomic state and avoided the transition to molecular phase.

The atomic hydrogen in the SMC also exhibits very similar properties. For instance, \citeauthor{dickey00} found that less than 15\% of the total \HI in the SMC is in the cold-phase. They also found the temperature of cold phase to be typically 40~K or less. 
They suggested that the difference in the properties of the CNM between the SMC and the Milky Way is a consequence of the difference in the metallicity in the ISM. Low-metallicity implies lower radiative cooling. In particular for clouds where cooling is dominated by fine-structure line emission, the thermal equilibrium between the warm and the cold phase requires higher pressures with decreasing metal abundance \citep{wolfire95}. This limits the existence of cold clouds to region of higher pressure, and hence the reduction in the cold gas fraction. 
Similar effects are observed in DLAs where the spin temperatures are inversely proportional to the metallicity of the DLA \citep{kanekar09}. Again, this emphasizes the inefficient cooling of gas at lower metallicities.

On the other hand, the lower metallicity would imply lower dust content and therefore, the fraction of atomic gas that is able to cool does not suffer from photoelectric heating by dust grains. However, the lack of dust grains is likely to impede the production of molecules. Dust plays a crucial role in shielding gas clouds against ionization thus aiding in atomic-to-molecular transition. Therefore, a drop in metallicity/dust content requires a much larger column of neutral gas to shield the clouds and form molecules \citep[see Fig.1 in ][]{krumholz09}. Observationally, this is evident from the variation in the minimum neutral gas column densities required to detect molecules in galaxies as a function of  metallicity. For example, in the Milky Way the transitions occurs when the neutral gas column density is $\rm10^{20.7}~cm^{-2}$ or higher \citep{savage77}. The same for LMC and SMC was found to be at N(HI)$\rm > 10^{21.3}~cm^{-2}$ and $\rm > 10^{22.0}~cm^{-2}$ respectively \citep{tumlinson02}.

Therefore, it is likely that UCG~7408 has a very low molecular gas fraction as compared to its neutral gas content. This would explain the small physical extent of the stellar component in this galaxy unlike its extended \HI distribution as well as the suppression in star formation (as indicated by the limiting star formation rate surface density in  Table~\ref{tbl2-sightlines}) at the position of the absorber despite its low temperature.

\subsection{Extended \HI Disks of  J141629 and J160659} 
  
We did not detect any \HI absorbers toward the sightlines 4 and 5 probing galaxy J141629 at impact parameters of 16.2 and 32.4~kpc and sightline 6 probing galaxy J160659 at an impact parameter of 23.3~kpc. The limiting optical depths and column densities estimated from 3$\sigma$ noise in the spectra are presented in Table~5. 

Both these galaxies have large \HI reservoirs ($\rm Log[M_{HI}] $= 9.4 and 9.9 respectively) and hence are expected to have extended \HI distributions. The empirical \HI\ mass to size relationship derived for dwarf galaxies from the WHISP Survey \citep{swaters02} predicts that the \HI distribution in J141629 and J160659 have radii, R$\rm _{HI}$, of 15.3 and 30.1~kpc, respectively. This implies that J160659 is probed by sightline \#6 ($\rho=23.3$~kpc) within the R$\rm _{HI}=30.1$~kpc. However, no \HI was observed in absorption. 
The 3$\sigma$ limit on the absorption column density towards this sightline is 3.6$\rm \times 10^{18}~cm^{-2}$. But, the expected \HI surface density within R$\rm _{HI}$ is $\geqslant \rm 1~M_{\odot}pc^{-2}$, which is equivalent to a column density of $\geqslant\rm 1.3~\times 10^{20}~cm^{-2}$. This implies that the limiting column density is 35 times lower than the column density predicted at R$\rm _{HI}$. 
Similarly, sightline \#5 probes J141629 at the edge of its \HI disk ($\rho=16.2$~kpc and $\rm R_{HI}=15.3$~kpc) and the upper limit on its \HI column places it an order of magnitude below that expected at $\rm R_{HI}$.
This suggest that cold-to-average \HI column densities could be lower than 10\% towards these sightlines.
A similar argument based on the empirically observed ratio between \HI to optical radius of $\sim$3 for dwarf galaxies also implies an unusually low cold \HI  (T$\sim$100~K or less) content at the position of these sightlines.

There could be two likely causes for the non-detection of \HI in these sightlines. The first possibility is that the \HI\ distribution is highly patchy and the covering fraction of gas with column density $\sim\rm 10^{20}~cm^{-2}$ is significantly smaller than 1. If that were the case, then it would also imply that true column density of the patchy \HI is much higher than 1.3$\rm \times 10^{20}~cm^{-2}$ within R$\rm _{HI}$. The second possibility is that the atomic gas is warm ($\rm T_{spin}>$1000~K) and produces a broad (due to Doppler broadening) and shallow  (i.e. low optical depth because $T_{spin} \propto (\tau)^{-2/3}$; see Eq. 5 from B11) absorption feature. Such features would remain hidden in our data due to the limited signal to noise ratio. This is a disadvantage that impacts most 21~cm absorption studies. 


\section{CONCLUSION   \label{Sec:conclusion}}

We presented follow-up VLA B-configuration observations of six quasar sightlines from the GBT 21cm \HI\ absorption sample of B11. The sightlines probed three foreground galaxies. This includes three sightlines probing dwarf galaxy, UGC~7408.
Strong \HI emission was detected towards these sightlines in the GBT spectra and hence a reliable limit on the optical depth based on the non-detection of absorption could not be made. Therefore, we obtained high spatial resolution imaging with the VLA and found the following:

\begin{itemize}
\item[1.] A narrow \HI absorber was detected towards sightline J122106.854+454852.16 (\#2) at 444.5~\kms. The absorber is associated with the extended \HI disk of UGC~7408 at an impact parameter of 2.8~kpc. This absorber was tentatively detected by B11 in their low spatial and spectral resolution VLA D-configuration data. Our higher resolution data showed the absorber to be narrow and unresolved. A Gaussian fit to our data gives a FWHM of 1.1~\kms and a corresponding kinetic temperature of  $\rm T_k \approx 26~K$. Assuming the kinetic temperature as a proxy for the spin temperature, we estimated the column density of the absorber to be $\rm 6.3\times 10^{19}~cm^{-2}$.
\item[2.] The area of the cloud associated with the absorber was measured to be $\rm 0.04~kpc^2$. This implies a covering fraction of the background source $\sim$25-30\%.
\item[3.] The cold-to-total \HI column density in the absorbing cloud towards this sightline was estimated to be $\sim$30\%. In terms of mass, the cold phase is about $\sim$10\% of the total \HI.
\item[4.] The low fraction of cold-to-total \HI and the existence of cold \HI at extremely low temperatures, where gas normally transitions to molecular phase, suggests that the process of condensation is suppressed in this galaxy. The likely explanation for the suppression is the low metal content in this galaxy. Similar properties of the CNM were seen in the SMC. Together, these two cases are consistent with the effects of low-metallicity in suppressing condensation.
\item[5.] We reported non-detection of \HI in absorption for the remaining five sightlines. The limiting  optical depths range from $\tau \le $0.01-0.39. These limits are about three times better than those measured by B11 towards most sightlines.
\item[6.] One of the sightlines is expected to probe the extended \HI\ disk ( $\rm \rho < R_{HI}$) of the foreground galaxy J160659.13+271642.6. The non-detection of \HI in absorption towards that sightline indicates an extremely low column density of cold gas ($\rm T_{spin}\lesssim 100~K$). Two likely interpretations are that either the cold gas distribution is highly patchy or the \HI is at much higher temperature ($\rm T_{spin}>$ 1000~K).
\item[7.] The non-detection of \HI puts stringent limits on \HI optical depth. This lends support to the conclusions from the B11 survey that the covering fraction of cold gas at column densities greater than that of DLA is very small beyond $\approx$15~kpc.
\end{itemize}

 Based on our results, we emphasize the need for a statistically larger sample to probe gas within 20~kpc of galaxies. 
Currently, plans are underway for two such surveys - the First Large Absorption Survey in HI \citep[FLASH; P.I. Sadler;][]{allison12} with the Australian Square Kilometre Array Pathfinder (ASKAP)  and the MeerKAT Absorption Line Survey (P.Is. Gupta \& Srianand) with the South African MeerKAT radio telescope.
Both of them are pathfinder missions for the Square Kilometer Array (SKA). The high sensitivity of SKA will make it the ideal telescope for mapping cold gas in absorption against fainter background sources.  These upcoming facilities and the planned surveys promise to substantially improve our understanding of small-scale properties of cold gas as a function of radial distance within their host galaxies. A large sample will also facilitate a census of variations in cold-gas properties and the process of condensation as a function of galaxy properties such as stellar mass, gas content, metallicity, SFR etc. That will be the crucial step in understanding the physics behind the process of condensation, its regulation, and how that translates to evolution of galaxies across the full stellar mass spectrum. \\

\noindent {\bf Acknowledgements}\\

This manuscript has benefited from insightful suggestions and comments from the referee.
We also thank M. Livio and L. Blitz for useful discussions.
We thank the staff of the Very Large Array for help during the data acquisition and reduction. 
SB acknowledges financial support from HST grant 12467\footnote{Support for program number 12467 was provided by NASA through a grant from the Space Telescope Science Institute, which is operated by the Association of Universities for Research in Astronomy, Inc., under NASA contract NAS5-26555.}  and from NRAO to cover travel and lodging during the data reduction under program 10C-120 at the  Pete V. Domenici Science Operations Center (SOC).
TMT appreciates support for this work from NSF grant AST­0908334.

Funding for the SDSS and SDSS-II has been provided by the Alfred P. Sloan Foundation, the Participating Institutions, the National Science Foundation, the U.S. Department of Energy, the National Aeronautics and Space Administration, the Japanese Monbukagakusho, the Max Planck Society, and the Higher Education Funding Council for England. 
The SDSS Web Site is http://www.sdss.org/. 
The SDSS is managed by the Astrophysical Research Consortium for the Participating Institutions. 
The Participating Institutions are the American Museum of Natural History, Astrophysical Institute Potsdam, University of Basel, University of Cambridge, Case Western Reserve University, University of Chicago, Drexel University, Fermilab, the Institute for Advanced Study, the Japan Participation Group, Johns Hopkins University, the Joint Institute for Nuclear Astrophysics, the Kavli Institute for Particle Astrophysics and Cosmology, the Korean Scientist Group, the Chinese Academy of Sciences (LAMOST), Los Alamos National Laboratory, the Max-Planck-Institute for Astronomy (MPIA), the Max-Planck-Institute for Astrophysics (MPA), New Mexico State University, Ohio State University, University of Pittsburgh, University of Portsmouth, Princeton University, the United States Naval Observatory, and the University of Washington.

\bibliographystyle{apj}	        
\bibliography{myref_bibtex}		

 \clearpage

\end{document}